\documentclass[a4paper,12pt]{article}
\usepackage{amsmath,amssymb,slashed}
\usepackage{jheppub}
\usepackage{xcolor,colortbl}
\usepackage{mathrsfs}
\usepackage{graphicx}
\usepackage{subcaption}
\usepackage{bbold}
\usepackage{physics}
\usepackage{booktabs} 
\usepackage{enumerate}
\usepackage{inputenc}
\usepackage{caption}
\usepackage{float}
\usepackage{hyperref}
\usepackage{multicol}
\usepackage{nicematrix}
\usepackage[normalem]{ulem}
\usepackage[compat=1.1.0]{tikz-feynman}
\usepackage{comment}
\usepackage{ragged2e}
\usepackage{etoolbox}

\title{Phasing out of Darkness: From Sterile Neutrino Dark Matter to Neutrino Masses via Time-Dependent Mixing}

\author[a]{Florian Goertz,}
\author[a]{Maya Hager,}
\author[b]{Giorgio Laverda,}
\author[c]{Javier Rubio}

\affiliation[a]{Max-Planck-Institut f\"ur Kernphysik, Saupfercheckweg 1, 69117 Heidelberg, Germany}
\affiliation[b]{Centro de Astrofísica e Gravitação  - CENTRA,
Departamento de Física, Instituto Superior Técnico - IST,
Universidade de Lisboa - UL, Av. Rovisco Pais 1, 1049-001 Lisboa, Portugal.}
\affiliation[c]{Departamento de Física Teórica and Instituto de Física de Partículas y del Cosmos (IPARCOS-UCM), Universidad Complutense de Madrid, 28040 
Madrid, Spain}

\emailAdd{florian.goertz@mpi-hd.mpg.de}
\emailAdd{maya.hager@mpi-hd.mpg.de}
\emailAdd{giorgio.laverda@tecnico.ulisboa.pt}
\emailAdd{javier.rubio@ucm.es}

\abstract{
Sterile neutrinos are a compelling candidate for generating neutrino masses and for elucidating the nature of dark matter. Astrophysical X-ray constraints on sterile neutrino dark matter decays, however, largely exclude the active-sterile mixing required to produce {\it simultaneously} the correct left-handed neutrino spectrum and keV-scale right-handed neutrino dark matter within a type-I seesaw framework. In this study, we demonstrate how these X-ray constraints can be circumvented through a time-dependent approach, thereby reviving a broad range of active-sterile mixing scenarios. Our minimal model incorporates two right-handed neutrinos, which form a two-component dark matter candidate, and an auxiliary scalar field that experiences a very late and still ongoing phase transition, leading to the spontaneous breaking of a global \( U(1)_N \) symmetry. Prior to this phase transition, only the right-handed neutrinos are massive, while the left-handed neutrinos remain massless because of the scalar field's vanishing expectation value. As the phase transition develops, the growing expectation value of the scalar field increases the active-sterile mixing, thereby opening dark matter decay channels and inducing neutrino masses. The time dependence allows the scenario to be consistent with X-ray constraints as well as current measurements of left-handed neutrino masses. The anticipated level of active-sterile mixing today is within the detection capabilities of the forthcoming TRISTAN (KATRIN) tritium-beta decay project. Additionally, cosmological surveys such as DESI or EUCLID and supernova neutrino observations can test the prediction of massless left-handed neutrinos prior to the phase transition.
}

\begin{document}
\maketitle
\flushbottom

\section{Introduction}

In the current paradigm of fundamental physics, neutrinos and dark matter (DM) represent two of the most solid arguments in favour of physics beyond the Standard Model (SM). On the one hand, the experimental observation of neutrino oscillations \cite{Kajita:2016cak,McDonald:2016ixn} indicates that despite the absence of right-handed (RH) neutrinos in the historical formulation of the SM, at least two of the three left-handed (LH) neutrinos have a (strikingly small) mass. On the other hand, the ubiquitous existence of DM is strongly suggested by both cosmological and astrophysical observations, while eluding direct and indirect detection so far \cite{Cirelli:2024ssz}. 

A fundamental connection between the above two major SM problems has been searched for since the seminal proposal of the seesaw mechanism for neutrino mass generation \cite{Minkowski:1977sc,Yanagida:1979as,Mohapatra:1979ia,Gell-Mann:1979vob,Schechter:1980gr}. In particular, the existence of a sterile neutrino $N$ with a large Majorana mass $M$ that mixes with the LH neutrinos leads naturally to a suppression of the active neutrino masses ${m_\nu \ll M}$, while fulfilling many DM requirements. However, with $\mathcal{O}(1)$ neutrino Yukawa couplings, the mass-range for sterile neutrinos to effectively suppress active neutrino masses is much too high to be a viable DM candidate, for which keV masses are favoured~\cite{Drewes:2016upu}. On top of the neutrino mass generation within the seesaw mechanism, the mixing of active and sterile neutrinos is the source of the radiative decay $N \rightarrow \nu \gamma$. Indeed, if DM consists of sterile neutrinos, DM dense regions are expected to emit observable X-ray photons of energy $E_\gamma = M/2$, with the non-observation of such a signal setting stringent constraints on keV sterile neutrino DM. In particular, requiring a level of active-sterile mixing below the X-ray observation is at odds with the generation of active-neutrino masses that can fit oscillation data. Therefore, seesaw-sterile neutrinos and DM-sterile neutrinos are generically assumed to be distinct particles with different origins. In this work, we propose an alternative picture in which the effective neutrino Yukawas arise from a recently started phase transition and the same keV sterile neutrinos can be DM and source the seesaw mechanism thanks to a time-dependent active-sterile mixing. 

A change with time of the neutrino masses has been considered in several different contexts, for instance in connection with dark energy and quintessence~\cite{Fardon:2003eh,Brookfield:2005bz,Wetterich:2007kr}, from a gravitational anomaly~\cite{Dvali:2016uhn} or through oscillating DM~\cite{Huang:2022wmz}, together with possible astrophysical and cosmological signatures \cite{Koksbang:2017rux, Lorenz:2018fzb, Lorenz:2021alz, deGouvea:2022dtw}. Changing Yukawas in the early universe have been studied e.g. in \cite{Jaramillo:2022mos}. In the present work, we introduce a mechanism for the generation of neutrino masses that takes place in the very recent cosmological history, namely after the Universe entered a dark-energy-dominated epoch. We consider a minimal extension of the SM including two $\mathcal{O}$(keV) RH neutrinos and an auxiliary scalar field responsible for the breaking of a new global $U(1)_N$ symmetry and leading to currently small but time-dependent effective neutrino Yukawas. Before the phase transition starts, only RH neutrinos are massive, whereas LH neutrinos remain massless because of a vanishing mixing. Both astrophysical X-ray constraints and cosmological bounds are fulfilled in this picture. As the transition unfolds, the increasing expectation value of the scalar field prompts a more efficient mixing, thus generating masses for the LH neutrinos, and opening DM decay channels. The crucial feature of this scenario lies in delaying the phase transition to the very late cosmological history. In particular, the scalar field might not have reached its vacuum expectation value by the present cosmological time, inducing a time-dependence in the active-sterile mixing and the neutrino masses. The active-sterile mixing becomes large enough today as to allow for the correct LH neutrino mass spectrum while remaining vanishingly small in the near past, thus lifting the constraints set by astrophysical X-ray observations. 

The main achievement of our setup is that sterile neutrinos can simultaneously play the role of DM while giving masses to the active neutrinos and avoiding the existing astrophysical constraints. A new window in the keV mass range becomes available for further probes, especially the proposed detector system TRISTAN in the KATRIN experiment \cite{KATRIN:2018oow} and, for the upper mass range, phase three of the HUNTER experiment \cite{Martoff:2021vxp}. Moreover, while the DESI (Dark Energy Spectroscopic Instrument) collaboration analysis of baryon acoustic oscillations \cite{DESI:2024mwx} is already pointing out to the possibility of having cosmologically massless neutrinos, the time-dependence of the active-sterile mixing can be further constrained by supernovae neutrino data, which would provide insights into neutrino oscillations at the time of emission \cite{deGouvea:2022dtw}.

This paper is organised as follows. In Section \ref{sec:neutrinoModel} we provide a comprehensive description of the general model setup. Section \ref{sec:scalar fieldPotAndPhTr} presents a model-independent analysis of the phase transition, detailing the requirements that any specific implementation must meet in order to open up previously excluded regions of the parameter space. Two such explicit realisations, based on non-minimal coupling to gravity and dynamical dark energy, respectively, are discussed in Appendix~\ref{app:ph_transition}. Section \ref{sec:constraints} addresses the constraints that neutrino oscillations and active-sterile mixing impose on the model parameter space, outlining also alternative probes for the model. Finally, Section \ref{sec:concl+outlook} offers a summary of our findings and provides an outlook on future research directions.

\section{The General Setup} \label{sec:neutrinoModel}

We focus on a minimal new-physics scenario involving only two RH neutrinos charged under an additional $U(1)_N$ global symmetry with quantum numbers $Q_{N}({\nu_{R}}_{1/2})=\mp 1$ and an auxiliary scalar field $S$ with charge $Q_{N}(S) = -1$ and eventually developing a non-zero expectation value $\bar S\equiv \langle S^c S \rangle^{1/2} \neq 0$. The terms leading to neutrino mass generation are given by\footnote{The mechanism employed here requires an even number of sterile neutrinos to explain DM, since for odd numbers the lightest sterile neutrino remains massless until very recently. Note also that, on general grounds, the number of scalar field insertions in the Majorana-like terms is determined by the sum of the individual charges of the RH flavour neutrinos $Q_N  ({\nu_R}_I)$, making the associated mass terms,
\begin{equation} \label{eq:Majorana}
\mathcal{L}\supset \frac{M_{IJ}}{2}  \left( \frac{ \bar S}{\Lambda_S} \right)^{\abs{Q_{N}({\nu_R}_I) + Q_{N}({\nu_R}_J)}} 
\overline{\nu_{R}^c}_I \, {\nu_{R}}_J\,,
\end{equation} 
strongly dependent on the expectation value of $S$ for arbitrary charge assignments. Since our main focus here is to explore the role of the sterile neutrinos as ${\cal O} ({\rm keV})$ mass DM candidates, we will restrict ourselves to cases where $Q_{N}({\nu_{R}}_I) + Q_{N}({\nu_R}_J)=0$ for some $I \neq J$, so no scalar field insertions are present in the relevant off-diagonal mass terms. This requirement is fulfilled in particular by the charge assignment $Q_N({\nu_R}_{1,2}) = \mp 1$ in Eq.~\eqref{eq:Yukawa}.}
\begin{eqnarray}    
    \label{eq:Yukawa}
    \mathcal{L} \supset && y_{\alpha 1} \frac{S^c}{\Lambda_S} \bar{L}_\alpha \,\tilde{H} \, \nu_{R1} + y_{\alpha 2} \frac{S}{\Lambda_S} \bar{L}_\alpha \,\tilde{H} \, \nu_{R2}  \nonumber \\
+ && \frac{M_{11}}{2} \left( \frac{S^c}{\Lambda_S} \right)^2 \overline{\nu_{R}^c}_1 \, {\nu_{R}}_1 + \frac{M_{22}}{2} \left( \frac{S}{\Lambda_S} \right)^2 \overline{\nu_{R}^c}_2 \, {\nu_{R}}_2 \\
+ && \frac{M_{12}}{2} \left(
\overline{\nu_{R}^c}_1 \, {\nu_R}_2 + \overline{\nu_{R}^c}_2 \, {\nu_R}_1 \right) + h.c. \nonumber 
\end{eqnarray} 
with $\Bar{L}$ the lepton doublet, $H$ the Higgs doublet ($\tilde{H} = i \sigma^2 H^\ast $), $\nu_R$ the RH neutrino, $\alpha = e, \mu, \tau$ denoting flavours, and $\Lambda_S$ a cutoff scale signalling the onset of new physics.\footnote{For the sake of generality, we remain agnostic about the specifics of an eventual UV completion. A natural scenario could involve, for instance,  a Dirac seesaw mechanism with heavy vector--like fermions. In this case, the cutoff scale in Eq.~\eqref{eq:Yukawa} would be dictated by the mass and couplings of the heavy fermions once these are integrated out \cite{CentellesChulia:2016rms,CentellesChulia:2018gwr}.}  All other SM components, including the lepton doublet, are taken to be uncharged under the new symmetry group, ensuring therefore that the usual electroweak symmetry breaking remains responsible for the masses of all the SM fields except neutrinos.  Previous works along these lines include, for instance, Refs.~\cite{Chikashige:1980ui,Gelmini:1980re,Georgi:1981pg,Chacko:2003dt},  where the full neutrino sector was taken to be charged under a $U(1)_L$ lepton number symmetry, or Refs.~\cite{Merle:2011yv,Barry:2011wb,Barry:2011fp}, which considered instead a Froggatt-Nielsen $U(1)_{FN}$ \cite{Froggatt:1978nt} in the neutrino sector. 

After the late symmetry breaking, masses are generated as 
\begin{equation}
    {\cal L} \supset \frac{1}{2} \left( 
    \begin{array}{cc}
        \overline{\nu_L} & \overline{\nu^c_R}  
    \end{array} 
    \right) \Tilde{M} 
    \left( 
    \begin{array}{c}
        \nu_L^c \\
        \nu_R
    \end{array} \right) + h.c,
\end{equation}
where the neutrino mass matrix associated to the Lagrangian density \eqref{eq:Yukawa} is given by
\begin{equation}
    \Tilde{M} = \left(
    \begin{array}{cc}
    0  & m_D \\
    m_D^T  & M_R \\
    \end{array}
    \right)\,,
\end{equation}
with 
\begin{equation}
    m_D =  \frac{v}{\sqrt{2}} \times \epsilon
    \left(
    \begin{array}{cc}
    y_{e1} & y_{e2} \\
    y_{\mu 1}& y_{\mu 2} \\
     y_{\tau 1}  & y_{\tau 2} 
    \end{array}
    \right) \equiv \frac{v}{\sqrt{2}} \times \epsilon \, Y \hspace{5mm}\textrm{and}    \hspace{5mm}  M_R = 
    \left(
    \begin{array}{cc}
    \epsilon^{2} \, M_{11} & M_{12} \\
    M_{12}  & \epsilon^{2} \, M_{22}  \\
    \end{array}
    \right)
    \label{eq:diracMajoranaMatrices}
\end{equation}
the Dirac and Majorana-like mass terms, $v=246\,$GeV the Higgs vacuum expectation value and $\epsilon \equiv \bar S /\Lambda_S$ a dimensionless parameter that will play an essential role in what follows. Note that the associated light neutrino mass matrix $M_\nu \approx - m_D M_R^{-1} m_D^T$ follows from a type-I seesaw structure, where a large part of the suppression originates from a small ratio $\epsilon \ll 1$ entering $m_D$, associated in our setting to a recently-started phase transition.

Before the phase transition occurs, the spectator field $S$ is locked at the origin of the effective potential (see \eqref{eq:scalarfieldPotential} below), the LH neutrinos are massless ($\epsilon=0$) and the two RH neutrinos are degenerate with mass $M^2 = M_{12}^2$. This situation changes at the onset of the transition, where the scalar field expectation value starts growing with time, inducing a non-vanishing mass for the LH neutrinos and an almost negligible mass splitting between the two sterile ones $(0 < \epsilon\ll 1)$. The resulting neutrino spectrum needs to be compatible with neutrino oscillations data and, in particular, with the two mass differences \cite{Esteban:2020cvm}
\begin{align}
    \Delta m_{21}^2 &= \left( 7.41_{-0.20}^{+0.21} \right) \times 10^{-5} \text{ eV}^2\,,  \nonumber \\
    \text{ (NO) } \Delta m_{31}^2 &= \left(2.511_{-0.027}^{+0.027}\right) \times 10^{-3} \text{ eV}^2 \nonumber \, ,\\ 
    \text{ (IO) } \Delta m_{32}^2 &= \left(- 2.498^{+0.032}_{-0.024} \right) \times 10^{-3} \text{ eV}^2 \,,
\end{align}
with normal ordering (NO) referring to a normal hierarchy where $\Delta m_{21}^2$ is the mass difference between the two lightest neutrinos and inverted ordering (IO) associated to a inverted hierarchy where $\Delta m_{21}^2$ stands for the mass difference between the two heaviest neutrinos. Since our scenario involves just two RH neutrinos, only two mass spectra are a priori possible, namely \newline
\begin{minipage}{.43\textwidth}
\raggedright
\begin{eqnarray}   
&\hspace{-8mm}\textbf{NO: }& \nonumber \\
    m_1 &=& 0 \text{ eV}\,,\nonumber \\
    m_2 &=& \sqrt{\Delta m_{21}^2} = 8.6 \times 10^{-3} \text{ eV}\,, \nonumber \\
     m_3 &=& \sqrt{\Delta m_{31}^2} = 5.0 \times 10^{-2} \text{ eV} \,,\nonumber 
\end{eqnarray}
\end{minipage}
\begin{minipage}{.43\textwidth}
\raggedright
\begin{eqnarray} \label{eq:massesIONO}
    & \hspace{-10mm} \textbf{IO: }&  \nonumber \\
    m_1 &=& \sqrt{\abs{\Delta m_{32}^2} - \Delta m_{21}^2} = 4.9  \times 10^{-2} \text{ eV}\,, \nonumber \\
    m_2 &=& \sqrt{\abs{\Delta m_{32}^2}}= 5.0  \times 10^{-2} \text{ eV}\,,\nonumber \\
    m_3 &=& 0 \text{ eV} \,. 
\end{eqnarray}
\end{minipage}
\newline \newline
\noindent In order to determine the current value of $\epsilon$ compatible with neutrino oscillations, we specify in what follows the late-symmetry-breaking dynamics. 

\section{Dynamics of Late Symmetry Breaking}
\label{sec:scalar fieldPotAndPhTr}

To parameterise the late-symmetry-breaking (LSB) pattern, we consider an effective potential for the $\bar S$ field,
\begin{equation}
    V(\bar S) = V_{0} + \frac12 \mu^2(t) \bar S^2 + V_{\rm HO}(\bar S) \,,
    \label{eq:scalarfieldPotential}
\end{equation}
with $\mu^2(t)$ a time-dependent mass term shifting from positive to constant negative values at a characteristic late symmetry-breaking time $t_{\rm LSB}$ and $V_{\rm HO}$ denoting potential higher-order operators, Higgs-portal interactions and $\Lambda_S$-suppressed interactions with neutrinos. In order to ensure that DM is sufficiently stable, we require the order parameter before the transition to exceed the sterile neutrino mass, i.e.~$\mu(t < t_{LSB}) > M$, such that the decay channel $N \rightarrow S + \nu_L$ is kinematically forbidden. On top of that, we assume the energy density of the $S$ field to stay completely subdominant during the whole cosmological history, reducing it to a mere spectator component with negligible backreaction effects on the background evolution. Finally, we consider the non-linear contributions $V_{\rm HO}$ to have a negligible effect on the scalar potential, such that no thermal corrections nor thermalisation effects are present in the scalar-field sector of the theory. 

Under these assumptions, the late-time cosmological history of the Universe can be well described by the usual Friedmann equation and the Klein-Gordon equation for the expectation value $\bar S$, namely
\begin{equation} \label{eq:hubble_function}
    H^2 = H_0^2 \left(\Omega_{\rm M} a^{-3} + \Omega_{\rm \Lambda} \right) \,, \hspace{10mm}  \Ddot{\bar S} + 3H\Dot{\bar S} + \mu^2(t) \bar S = 0 \,,
\end{equation}
with $H=\Dot{a}/a$ the Hubble rate and $ \Omega_{\rm M}$ and $ \Omega_{\rm \Lambda}$ the present-day matter and dark energy densities. Assuming the phase transition to proceed fast enough as compared to the timescale $\Delta t=t-t_{\rm LSB}$, this system of equations admits an approximate solution 
\begin{equation}
    a(t)=\left( \frac{\Omega_{\rm M}}{\Omega_{\Lambda }} \right)^{1/3} \sinh^{2/3} \left( \frac{3\sqrt{\Omega_{\Lambda }}H_0 \, t}{2} \right)\,, \hspace{6mm}
     \bar S(t) \simeq \, \bar S_0\left(\frac{a(t_{\rm LSB})}{ a(t) } \right)^{3/2}\exp\left[{\Bar{\mu}\Delta t }\right]
    \,, 
    \label{eq:expect_val_approx}
\end{equation}
at $t>t_{\rm LSB}$, with $\Bar{\mu}\equiv \abs{\mu(t>t_{LSB})}$ and the integration constant 
\begin{equation}
   \bar  S_0 =  \frac{H(t_{\rm LSB}) \, }{2}\left[1 - \frac{5 H(t_{\rm LSB})}{2 \Bar{\mu}}
    \right]
\end{equation}
following from imposing typical quantum fluctuations in an expanding background as initial conditions, $\bar S(t_{\rm LSB}) \sim H(t_{\rm LSB})$ and $\dot{\Bar{S}}(t_{\rm LSB}) \sim H^2(t_{\rm LSB})$. Note that, as long as the timescale associated with the time-changing mass $\mu(t)$ is much shorter than the evolution timescale of $\bar S(t)$, this behaviour is completely independent of the exact mechanism responsible for the generation of the tachyonic mass. For instance, if the sudden transition was replaced by a hyperbolic-tangent parametrisation ${\mu^2(t)=\mu^2_0 - \alpha^2\tanh ( \beta(t-t_0))}$ smoothly interpolating between asymptotic states $\mu^2_{t\to-\infty}=\mu^2_0+\alpha^2$ and $\mu^2_{t\to\infty}=-\Bar{\mu}^2=\mu^2_0-\alpha^2$, the speed of the transition would have to satisfy $\tau^{-1}=\beta \gg \Bar{\mu}$. Some specific implementations of the phase transition are presented in Appendix \ref{app:ph_transition}. 

\section{Experimental and Observational Constraints}\label{sec:constraints}

Having specified the LSB dynamics, we proceed now to determine the current value of $\epsilon$ and the active-sterile mixing needed to reproduce the mass spectra in \eqref{eq:massesIONO}, discussing also the subsequent evolution of the model parameters and the impact of the LSB on current constraints and potential further probes.

\subsection{Neutrino Mixing and Masses}\label{sec:LHmassesMixings} 

To determine the magnitude of the expectation value $\bar S$ needed to reproduce the correct level of mixing and spectra of LH neutrinos today, we perform a numerical scanning of the parameter space using a Casas-Ibarra parametrisation \cite{Casas:2001sr, Ibarra:2005qi} and the best fit values for masses, mixings, and CP violation in Ref.~\cite{Esteban:2020cvm}. Since the Majorana-like matrix in our model is not diagonal, the first step is to bring it to a diagonal form $d_M$. To this end, we consider an initial unitary transformation $d_M = U_R^T \, M_R \, U_R $, with
\begin{equation}
    U_R = \left( \begin{array}{cc}
        i \cos \theta & \sin \theta \\
        - i \sin \theta & \cos \theta 
    \end{array} \right)
\end{equation}
and $\theta = \pi / 4 + \mathcal{O}(\epsilon^2) \approx \pi/4$. Inverting this relation,  $M_R^{-1} = U_R \, d_M^{-1} \, U_R^T$, replacing the result into the type-I seesaw formula $  M_\nu = - m_D M_R^{-1} m_D^T$ and diagonalising the obtained active neutrino mass matrix with  the usual Pontecorvo–Maki–Nakagawa–Sakata matrix $U$ \cite{Maki:1962mu}, we obtain $d_m = U^T M_\nu  U$. In terms of known parameters, this translates into a Yukawa matrix\footnote{The definition of the Yukawa matrix in Ref.~\cite{Ibarra:2005qi} is the transpose of $Y$.}
\begin{equation}
    Y = \frac{\sqrt 2}{\epsilon v} \left( U_R^\ast \sqrt{d_M} R \sqrt{d_m} U^\dagger \right)^T \,,
    \label{eq:YukCIpar}
\end{equation}
with 
\begin{equation}
    R_{NO} = \left( \begin{array}{ccc}
        0 & \cos z & \zeta \sin z \\
        0 & - \sin z & \zeta \cos z
    \end{array}\right)\,, \quad \quad \quad \quad
    R_{IO} = \left( \begin{array}{ccc}
        \cos z & \zeta \sin z & 0 \\
        - \sin z & \zeta \cos z & 0
    \end{array}\right)\,,
    \label{eq:Rmatrices}
\end{equation}
$\cos z = \cos (x + i y)$ and $\zeta = \pm 1$, accounting for further freedom in parameters, not fixed by the observed masses and mixings.  For the sterile neutrino masses in $d_M$, we vary the entries $M_{11},M_{22},M_{12}$ in $M_R$ within $[0.4,50]$ keV, scanning also the $R$ matrices with random values $\zeta = \pm 1$ and $\{x,y\}$ coordinates in the range $x \in [0,2 \pi]$ and $y \in [-10, 10]$.~\footnote{Since $\cos (x + i y) = \cos x \cosh y - i \sin x \sinh y$ and $\sin (x + i y) = \cosh y \sin x + i \cos x \sinh y$, there exists a clear periodicity in $x$, while $y$ can take arbitrary real values. Nonetheless, since extreme values for $y$ lead to extreme absolute values in the Yukawa matrix $Y$ which are eventually filtered out by our perturbativity constraint, we restrict ourselves to a range $y \in [-10,10]$.} The lower cut ($0.4$ keV) imposed on the DM mass stems from phase space arguments for fermionic DM, i.e. the Tremaine-Gunn bound \cite{Tremaine:1979we}, whereas the upper bound ($50$ keV) is conservatively chosen to account for several potential DM production mechanisms. 
In particular, we remain agnostic about the specific mechanism responsible for the production of sterile neutrinos, assuming only that the correct DM abundance is generated non-thermally. This assumption avoids potential issues with big bang nucleosynthesis (BBN) that could arise from new thermalized light degrees of freedom. For instance, sterile neutrinos could be produced as decay products of the inflaton field or during a period of kination, a phase naturally occurring in quintessential inflation scenarios. In these cases, due to the negligible interactions of the sterile neutrinos in the early universe, they would not thermalize. For a comprehensive review of sterile neutrino dark matter and various production mechanisms, see \cite{Boyarsky:2018tvu}.

Scanning over 3000 sets of mass matrices $M_R$, finding for each of them 30 parameter combinations of $\epsilon$ and $Y$ able to reproduce the correct spectra \eqref{eq:massesIONO} and discarding Yukawa matrix elements $Y_{ij}$ with absolute magnitudes $10<\abs{Y_{ij}}$ and $\abs{Y_{ij}}<10^{-2}$ for the sake of perturbativity and naturalness, we find that a ratio
\begin{equation}
    \epsilon \, (t_0) = \frac{\bar S(t_0)}{\Lambda_S} \sim 10^{-11} - 10^{-9}
    \label{eq:vevRatio}
\end{equation}
correctly reproduces the allowed LH neutrino masses while providing also a viable $\cal{O} (\rm{keV})$ sterile neutrino DM candidate at the present cosmological time $t_0$\footnote{Our mechanism can be extended to higher sterile neutrino mass ranges, although this comes with an associated increase in the required active-sterile mixing today, due to the additional suppression of active neutrino masses. Achieving this increased mixing can be realized in two ways: either the phase transition occurs earlier, or the broken potential becomes steeper while maintaining the symmetry unbroken until very late times. The former approach complicates the task of alleviating astrophysical constraints, while the latter is less attractive from a model-building perspective, as it relies heavily on fine-tuning the phase transition mechanism. Still, it would be interesting to explore in future work how, for instance, gamma-ray constraints on GeV-scale sterile neutrino dark matter might be evaded.}. As explicit in the latest expression, there exists an intrinsic degeneracy between the scalar field expectation value  and the scale of new physics, being always possible to  increase/decrease the needed value of $\bar S$ in a specific LSB realisation by increasing/decreasing $\Lambda_S$.~\footnote{By increasing the $U(1)_N$ charge of the RH neutrinos, one could achieve an additional suppression, as is done in the Froggatt-Nielsen mechanism \cite{Froggatt:1978nt}, and still increase $\bar S (t_0)$ further.}  Conservative lower and upper bounds on the scalar field expectation value follow, however, from requiring the absence of sizeable fifth forces ($\Lambda_S \gtrsim 10^4$\,GeV, cf. Appendix \ref{app:fifthforce}) and strictly sub-Planckian cutoff scales ($\Lambda_S < M_{P} \sim 10^{19}$ GeV),
\begin{equation}\label{Srange}
\bar S(t_0) \sim \left( 10^{-7} - 10^{10} \right) \text{GeV}\,.
\end{equation}

\begin{figure}[tb]
\centering
\includegraphics[width=.8\textwidth]{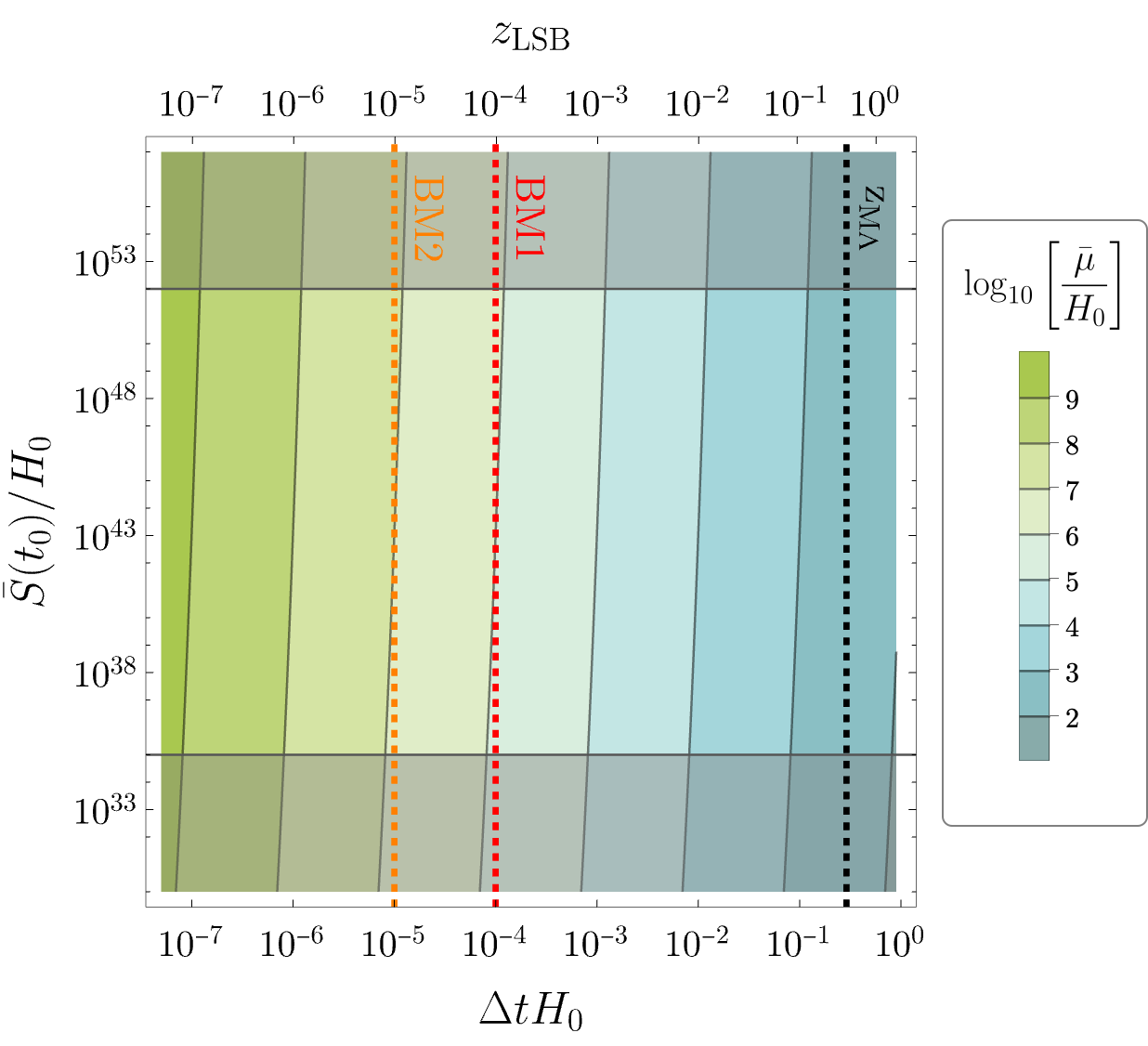}
\caption{Expectation value of the scalar field $\bar S(t_0)$ at the present cosmological time $t_0$ as a function of the symmetry-breaking timescale $\Delta t = t_0 - t_{\rm LSB}$ (or redshift $z_{\rm LSB}$) and the tachyonic mass $\Bar{\mu}$ as in~\eqref{eq:expect_val_approx}. The shaded areas are excluded by the constraint \eqref{Srange} on the effective neutrino Yukawas. The black dashed vertical line indicates the matter-dark energy equality time while the coloured dashed lines highlight the two benchmark scenario explored in Section \ref{sec:activeSterileMixing}.} \label{fig:scalar_field_evolution}
\end{figure}

The present-time expectation value computed from~\eqref{eq:expect_val_approx} is displayed in Figure~\ref{fig:scalar_field_evolution} as a function of the symmetry breaking time $\Delta t = t_0 - t_{\rm LSB}$ and the tachyonic mass $\Bar{\mu}$, with the areas excluded by \eqref{Srange} shaded in grey.
In the allowed parameter space, two different scenarios can occur. If the scalar field possesses a small tachyonic mass $\Bar{\mu} \lesssim 10^3 H_0$, its symmetry breaking must happen relatively early in the cosmological history (redshift $z \gtrsim 0.3$), such that a large expectation value is achieved at present time. Alternatively, thanks to a large tachyonic mass $\Bar{\mu} \gtrsim 10^3 H_0$, the phase transition can happen very recently, with a fast exponential growth. Indeed, the rate of change of the expectation value in \eqref{eq:expect_val_approx} is proportional to the field's mass, $
\dot{\bar S}/\bar S\sim \Bar{\mu}$, 
and therefore, a later phase transition is always characterised by a steeper tachyonic growth. As a self-consistency check, we note also that the tachyonic scalar field remains always a subdominant component during the considered cosmological history. In particular, for $\Bar{\mu} \leq 10^7\times H_0$ the present-day $\bar S$ energy density $\Delta \rho_S \sim \Bar{\mu}^2 \bar S^2(t_0) \in \Bar{\mu}^2 \times [10^{70}-10^{100}] \times H^2_0$ is many orders of magnitude smaller than the dark-energy counterpart $\rho_{\Lambda} \sim 3 M_{P}^2 H_0^2 \sim 10^{120} \times H^4_0$. The expectation value is expected to increase further, unless the field is stabilised by quartic or higher-order operators.

\subsection{Active-Sterile Mixing}\label{sec:activeSterileMixing}

\begin{figure}[ht!]
\centering
    \begin{subfigure}{0.81\textwidth}
    \includegraphics[width=\linewidth]{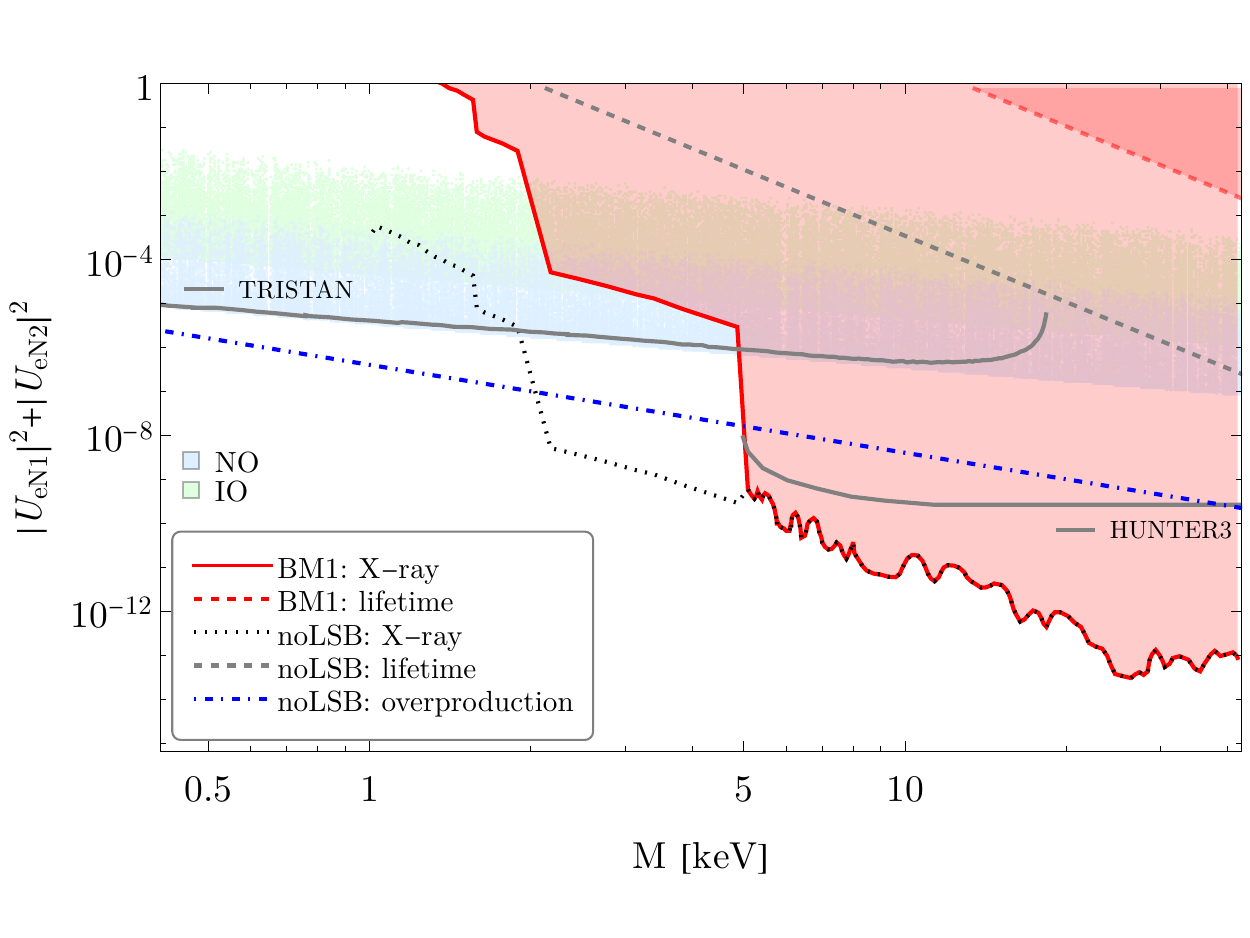}\vspace{-6.5mm}
    \end{subfigure}
    \begin{subfigure}{0.81\textwidth}
    \includegraphics[width=\linewidth]{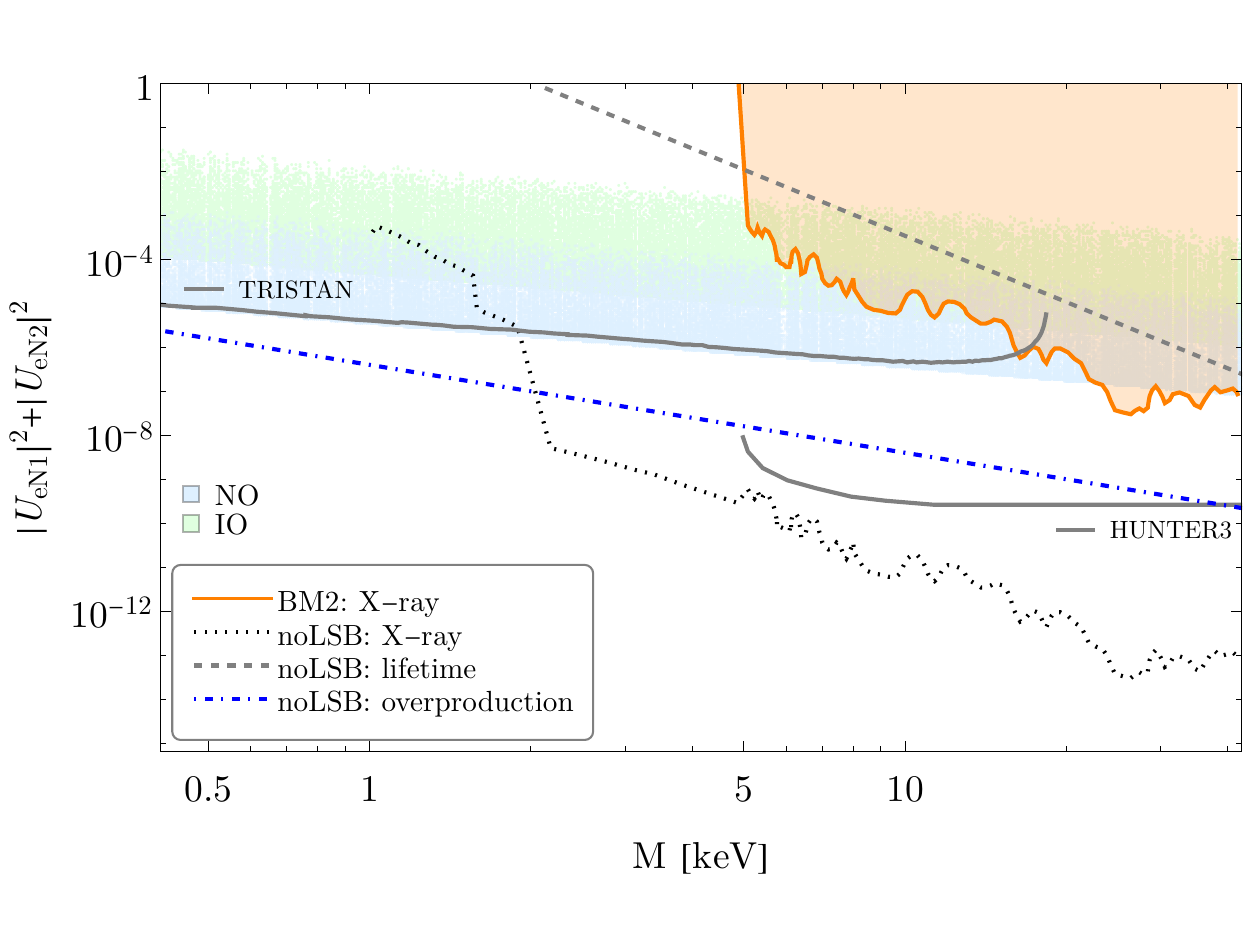}
    \end{subfigure}\vspace{-5mm}
    \caption{Active-sterile mixing $\abs{U_{eN1}}^2 + \abs{U_{eN2}}^2$ versus sterile neutrino mass $M$ for both normal ordering (NO, light blue) and inverted ordering (IO, light green). See text for details on how the constraints obtained without assuming a late phase transition (noLSB), excluding the regions above the dashed/dotted/dotted-dashed lines, are modified in the benchmark models BM1 (upper panel) and BM2 (lower panel). If sterile neutrinos are a thermal relic, Lyman-$\alpha$ constraints require ${M > 1.9 \text{ keV}}$ \cite{Garzilli:2019qki}. The grey lines show the future sensitivity of the TRISTAN detector system for KATRIN \cite{KATRIN:2018oow} and of the proposed HUNTER experiment \cite{Martoff:2021vxp}.}
    \label{fig:XRayBounds}
\end{figure}

The most stringent constraints on sterile neutrino DM stem from the effective mixing $U_{\alpha I}\sim (y \, \epsilon \, v)/M$ between sterile and active neutrinos induced by the late phase transition, namely
\begin{equation}
    \nu_{L\alpha} = \sum_{i=1}^3 U_{\alpha i} \nu_i + \sum_{I=1}^2 U_{\alpha N_I} N_I^c\,,
\end{equation}
with the indices $i$ and $I$ summing over active neutrino mass eigenstates and sterile neutrinos, respectively, and $\alpha=e,\mu,\tau$ denoting flavour. The values of $U_{\alpha N_I}$ follow immediately from the viable parameter space obtained in Section \ref{sec:LHmassesMixings}. In particular, as seen in the light blue and light green regions in Fig.~\ref{fig:XRayBounds}, a normal hierarchy (NO) with one zero-mass neutrino features generically lower mixing angles than an inverse hierarchy (IO), as the sum of active neutrino masses is lower, therefore requiring less mixing.

The main bounds on the active-sterile mixing are related to the potential DM overproduction, the DM decay into active neutrinos $N \rightarrow 3 \nu$ and the radiative decay $N \rightarrow \nu + \gamma$, cf.~Fig.~\ref{fig:decaysAfterPhTr}. 
The first of these constraints, excluding the parameter space above the blue dotted-dashed line in Fig.~\ref{fig:XRayBounds}, is automatically avoided in our scenario by means of the suppressed coupling between LH and RH neutrinos in the early Universe, which prevents the overproduction of the latter via active-sterile mixing.~\footnote{The constraint for the overproduction is indeed model-dependent, but, if active and sterile neutrinos mixed in the early Universe, sterile neutrinos would be produced via scattering-induced decoherence, even if the main production mechanisms to obtain the correct DM relic abundance were to differ \cite{Merle:2015vzu}.} To determine the restrictions imposed by the remaining decay channels on the parameter space of the theory, we consider two benchmark scenarios where the LSB would have started $10^6$ years ago (BM1) and only $10^5$ years ago (BM2): 
\begin{itemize}
\item {\bf DM Decay into Neutrinos:} The decay width for $N \rightarrow 3 \nu$ is given by \cite{Lee:1977tib,Pal:1981rm,Barger:1995ty}
\begin{equation}
    \Gamma_{N \rightarrow 3 \nu} = \frac{G_F^2 M^5}{96 \pi^3} \sin^2 \theta = \frac{1}{1.5 \times 10^{14} \text{ s}} \left( \frac{M}{10 \text{ keV}} \right)^5 \sin^2 \theta\,,
    \label{eq:3nudecaywidth}
\end{equation}
with $M$ the sterile neutrino mass and $\sin \theta = \abs{U_{\alpha N_I}}$. In the absence of a late phase transition (noLSB) and for the $\mathcal{O}( \rm{keV})$ DM masses under consideration, this would correspond to a DM lifetime $\tau_{\rm DM}=\Gamma_{\rm DM}^{-1}$ significantly shorter than the age of the Universe, $\tau_{\text{U}}=10^{10}$ yrs, unless the  mixing angle $\theta$ is properly tuned to 
\begin{equation}
    \theta^2 < 3.4 \times 10^{-4} \left( \frac{10 \text{ keV}}{M} \right)^5\,. \quad\quad\quad \quad\text{ (noLSB)}
\end{equation}
In the present model, however, the sterile neutrino DM candidate starts mixing with the LH neutrinos only very recently, effectively shortening the required DM lifetime to $\tau_{\rm DM} > \tau_{\text{U}} - t_{\text{LSB}}$. This translates into a mixing angle constraint which is at least four to five orders of magnitude less stringent, namely ${\theta^2 < 3.4 \times (10 \text{ keV}/M)^5}$ for BM1 and ${\theta^2 < 34 \times (10 \text{ keV}/M)^5}$ for BM2.  
\begin{figure}
\centering
\[
\begin{aligned}
\begin{tikzpicture}
  \begin{feynman}
    \vertex [crossed dot] (i1) {} ;
    \vertex [left =of i1] (i3) {};
    \vertex [right =of i1, dot] (i4) {};
    \vertex [below right =of i4] (i2) {};
    \vertex[above right =of i4, dot] (i5) {};
    \vertex[below right =of i5] (i6) {};
    \vertex[above right =of i5] (i7) {};
    \diagram* {
       (i3) -- [fermion] (i1)-- [fermion,edge label=\(\nu_e\)] (i4);
       (i4) -- [fermion] (i2) ;
       (i4) -- [boson,edge label=\(Z\)](i5) -- [fermion](i6);
       (i7) -- [fermion] (i5);
    };
    \vertex [right=0.5em of i7] {\(\Bar{\nu}_\alpha\)};
    \vertex [right=0.5em of i6] {\(\nu_\alpha\)};
    \vertex [right=0.5em of i2] {\(\nu_e\)};
    \vertex [left=0.5em of i3] {\(N\)};
    \vertex [above =1.2em of i1] {\(U_{eN}\)}; 
  \end{feynman}
\end{tikzpicture}
\end{aligned}
\begin{aligned}
\begin{tikzpicture}
  \begin{feynman}
    \vertex [crossed dot] (i1) {\( \)} ;
    \vertex[right =of i1, dot] (i2) {};
    \vertex[left =of i1] (i3) {};
    \vertex[right =of i2] (i4p) {};
    \vertex[above right =of i2, dot] (i4pp) {};
    \vertex[below right =of i4pp, dot] (i4) {};
    \vertex[right =of i4] (i5) {};
    \vertex[above right =of i4pp] (i6) {};

    \diagram* {
       (i3) -- [fermion, edge label=\(N\)] (i1) -- [fermion] (i2) -- [boson, edge label=\(W^{\pm}\)] (i4) -- [fermion, edge label=\(\nu_e\)] (i5);
       (i2) -- [fermion, quarter left, edge label=\(e^{\mp}\)] (i4pp) -- [fermion, quarter left] (i4);
       (i4pp) -- [boson, bend left] (i6);
    };
    \vertex [above =1.2em of i1] {\(U_{eN}\)};
    \vertex [right =0.5em of i6] {\(\gamma\)};
  \end{feynman}
\end{tikzpicture}
\end{aligned}
\]
\caption{Sterile neutrino decays: $N \rightarrow 3 \nu$ (left), $N \rightarrow \nu + \gamma$ (right).}
\label{fig:decaysAfterPhTr}
\end{figure}
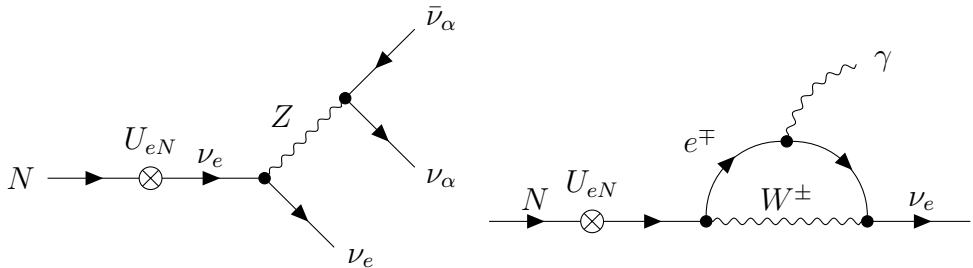
\noindent Note that these values should be understood as rather conservative estimates, as they are based on the assumption of an instantaneous transition of $\epsilon$ from being vanishing to its value today. In realistic LSB embeddings, however, the neutrino mixing is expected to change in a continuous manner, leading to a further relaxation of the above constraints. 

\item {\bf Radiative DM Decay:} The decay rate for $N \rightarrow \nu+\gamma $ is given by \cite{Shrock:1982sc,Fermi-LAT:2012ugx,Dugger:2010ys,Garny:2012vt}
\begin{equation}
    \Gamma_{N \rightarrow \gamma \nu} = \frac{9 \alpha G_F^2}{256 \times 4 \pi^4} \sin^2 2 \theta M^5 = 5.5 \times 10^{-22} \theta^2 \left( \frac{M}{1 \text{ keV}} \right)^5 \frac{1}{\text{s}}\,.
    \label{eq:xraydecaywidth}
\end{equation}
Although this width is significantly smaller than \eqref{eq:3nudecaywidth} and would naively give rise to sufficiently long-lived DM candidates, it may become boosted in sufficiently dense DM regions, producing photons with energies within the reach of X-ray telescopes, $E_\gamma \approx M/2$. The most stringent limits on active-sterile mixing come indeed from the non-observation of such a monochromatic X-ray signal from sources at a wide range of distances from Earth. The currently available constraints are summarised in Table \ref{tab:Xray}, with the upper part listing X-ray analyses of structures more than $10^5$ lyr away and the lower part referring to sources within our galactic centre and bulge, i.e. merely $10^4$ lyr from Earth.

\begin{table}
    \centering
    \begin{tabular}{l|l|c}
\toprule\toprule 
X-ray observation & Validity & Impact of LSB \\
\midrule\midrule 
 Dwarf Ursa Minor  \cite{Loewenstein:2008yi}  &  $d \sim $ 2.3 $\times 10^5$ lyr & \\
Dwarf Draco \cite{Riemer-Sorensen:2009zil}  &  $d \sim 2.6 \times 10^5$ lyr & \\
Dwarf satellite galaxies + M31   \cite{Horiuchi:2013noa}  &  $d \gtrsim $ few $\times\,  10^5$ lyr & \\
Dwarf spheroidal galaxies \cite{Malyshev:2014xqa} &  $d \gtrsim $ few $\times \,10^5$ lyr & \\
Galaxy clusters  \cite{Iakubovskyi:2015dna} &  $d \gtrsim $ 5 $\times \, 10^5$ lyr \\
M31 \cite{Watson:2006qb,Ng:2019gch} &  $d \sim 2.5 \times 10^6$ lyr & Softened for  \\
Coma \& Virgo cluster    \cite{Boyarsky:2006zi} & $d \gtrsim $ few $\times 10^7$ lyr & BM1 and BM2 \\ 
Perseus cluster  \cite{Tamura:2014mta,Tamura:2018scp} & $d \sim 2.4 \times 10^8$ lyr \\
Bullet cluster \cite{Riemer-Sorensen:2015kqa}  & $d \sim 3.7 \times 10^9$ lyr & \\        
 X-Ray Background \cite{McCammon:2002gb,Boyarsky:2005us,Boyarsky:2006ag,Boyarsky:2006hr, Sekiya:2015jsa} &  See comment \cite{clarification} & \\
Cosmic X-Ray Background \cite{Abazajian:2006jc,Riemer-Sorensen:2006pdg} & See comment \cite{clarification} \\ [0.2cm]
\midrule 
Milky Way centre \cite{Foster:2021ngm}& $M = 5 - 16$ keV & \\
Milky Way     \cite{Neronov:2016wdd,Krivonos:2024yvm}  & $M = 6 - 40$ keV &  \\
Milky Way bulge \cite{Perez:2016tcq}& $M = 6 - 40$ keV &  Unchanged for BM1 \\
Milky Way bulge  \cite{Roach:2019ctw}  & $M = 10 - 40$ keV & Softened for BM2 \\
Milky Way centre \& halo  \cite{Ng:2015gfa}  & $M = 20 - 50$ keV & \\
Milky Way \cite{Boyarsky:2007ge,Yuksel:2007xh} & 40 keV $\leq M$ \\ [0.15cm]
\bottomrule\bottomrule
    \end{tabular}
    \caption{Overview of X-ray limits on sterile neutrino DM. In addition to the constraints stemming from specific objects, we consider measurements of the X-ray background and the cosmic X-ray background. The constraints in the upper part of the table can be alleviated by our benchmark scenarios BM1 and BM2, while the limits in the lower part remain unchanged in BM1, but can be softened by BM2.}
    \label{tab:Xray}
\end{table}

In the model presented here, since the phase transition is ongoing and the scalar field is continuously rolling down its potential, the mixing angle at photon production is significantly smaller than today. Taking into account that X-rays need to travel towards the Earth before being detected, this translates into a weakening of the associated constraints as compared to those for models without a late phase transition.  The degree to which the constraints are alleviated depends on the distance between the source of emission and the Earth. Another possibility to evade X-ray bounds can come for instance through a reduction in the branching ratio $N \rightarrow \nu \gamma$ via a cancellation with a new physics diagram \cite{Benso:2019jog}.

In BM1, from the time of transition $\Delta t H_0=10^{-4}$ the scalar field expectation value $\bar S$ increased by two orders of magnitude within the last $10^5$ years. Taking into account that $U_{\alpha I} \sim \epsilon$, the squared mixing value at production is at least four orders of magnitude smaller than today, being even smaller before $10^5$ years ago. However, photons from the galactic centre of the Milky Way (MW) that are observed today originated only $10^4$ years ago, and thus did not experience a significant change in $\epsilon$ on that timescale. The subset of constraints arising from observations of the MW centre are thus not altered and only the parameter space affected by the constraints in the upper portion of Table \ref{tab:Xray} opens up for $M \lesssim 5$ keV. On the other hand, for BM2 the transition happens even more recently at $\Delta t H_0=10^{-5}$ and $\epsilon$ increases by three orders of magnitude in the last $10^4$ years, modifying the mixing squared by six orders of magnitude and ameliorating all constraints. 
\end{itemize}

The above discussion is summarised in Fig.~\ref{fig:XRayBounds}, where we display the full parameter space satisfying all the requirements on neutrino masses and mixings for both NO and IO.~\footnote{Following Ref.~\cite{Drewes:2016upu}, the X-ray limits are divided by a factor of two in order to account for uncertainties in the DM content of the corresponding structure.} Since our model features two RH neutrinos with almost degenerate masses, the constraints apply to the sum of squared active-sterile mixings $\abs{U_{eN1}}^2 + \abs{U_{eN2}}^2$. As apparent in the upper and lower plot, the mixing required for reproducing the correct neutrino spectrum in the presence of a LSB is no longer at odds with X-ray bounds. In particular, $\mathcal{O}$(keV) sterile neutrino DM can source the observed neutrino masses without being observationally excluded. Furthermore, the predicted parameter space for both NO and IO is testable. Indeed, the lifting of the X-ray bounds in our model opens up parameter space for sterile neutrino DM that will be probed by the KATRIN extension TRISTAN via tritium-beta decay \cite{KATRIN:2018oow} in a region that was previously thought to be excluded for keV sterile neutrino DM (unless a model enhances the contribution to beta decay while keeping the mixing small, see \cite{Barry:2014ika}). The grey line in Fig.~\ref{fig:XRayBounds} shows the future design sensitivity of TRISTAN. The statistical limit after three years of data taking with the full source strength of the KATRIN experiment will reach even lower. An additional upcoming experiment by the Heavy Unseen Neutrinos from Total Energy-momentum Reconstruction (HUNTER) collaboration is based on radioactive atom trapping and high-resolution decay-product spectrometry \cite{Martoff:2021vxp}. The sensitivity window reached by HUNTER phase 3 (grey line labelled ''HUNTER3") will no longer be in conflict with X-ray constraints if the LSB happens sufficiently recent, as is the case in BM2.

\subsection{Additional Probes}

There exists a multitude of probes for sterile neutrino properties and neutrino masses. In the subsequent list, we describe to what extent they apply to the scenario considered here, illustrating also additional predictions of the model.

\begin{itemize}
    \item {\bf Neutrino Masses from Cosmology:}
    The combination of Cosmic Microwave Background observations \cite{Planck:2019nip,Carron:2022eyg,ACT:2023dou,ACT:2023kun} and the analysis of baryon acoustic oscillation from the first year of data taking of the Dark Energy Spectroscopic Instrument (DESI) \cite{DESI:2024mwx} has set a stringent limit ${\sum m_\nu < 7 \times 10^{-2} \text{ eV}}$ on the sum of neutrino masses, also consistent with an earlier (e)BOSS constraint $\sum m_\nu < 8.2\times 10^{-2}$ eV~\cite{Brieden:2022lsd}. On top of excluding the minimum sum in IO at approximately $3\sigma$, the posterior distribution is peaked around ${\sum m_\nu \simeq 0}$, being close to exclude also the minimum value allowed by oscillation experiments ${\sum m_\nu \geq 5.9 \times 10^{-2} \text{ eV}}$ at $2\sigma$ confidence level. Although problematic for standard new-physics models not involving a time-dependent sterile neutrino mixing, this result is in excellent agreement with the scenario under consideration, where the active neutrino species remain massless throughout most of the evolution of the Universe. Indeed, we expect future cosmological bounds to exclude the minimal sum of neutrino masses expected from the oscillation data.
    
    \item {\bf Neutrino Masses from KATRIN:}  Neither the current upper bound on neutrino masses from the KATRIN experiment, $m_\nu < 0.8$ eV (90 \% CL) \cite{KATRIN:2021uub}, nor the experiment's design limit of $0.2$ eV \cite{KATRIN:2005fny} are sensitive to the sum determined from oscillations while one neutrino is massless. However, potential extensions of the minimal scenario involving, for instance, four RH neutrinos, with the two lower-mass ones playing the role of DM, could lead to an absolute neutrino mass scale close to the current KATRIN reach, while continuing to avoid cosmological limits. A dedicated study with four RH neutrinos is left for future work.
    
    \item {\bf Supernovae Neutrinos:} For very recent phase transitions, the neutrinos emitted at sufficiently far away supernovae (SN) will not have been able to oscillate inside the star, gaining masses only while travelling towards Earth. Ref.~\cite{deGouvea:2022dtw} map how the probing of relic neutrinos from the diffuse supernovae neutrino background (DSNB) can search for late-time neutrino mass generation. Although Super-Kamiokande (SK) is already anticipating finding the DSNB within the next 10 years \cite{Super-Kamiokande:2013ufi,Super-Kamiokande:2021jaq}, additional studies will come from future experiments such as Hyper-Kamiokande (HK) \cite{Hyper-Kamiokande:2018ofw}, the Juangmen Underground Neutrino Observatory (JUNO) \cite{JUNO:2015zny}, and the Deep Underground Neutrino Experiment (DUNE) \cite{DUNE:2020ypp}. Furthermore, the total DSNB flux will be probable via coherent neutrino-nucleus scattering (CE$\nu$NS), see \cite{Pattavina:2020cqc,Suliga:2021hek,Baum:2022wfc}. Although in 1987 a small number of SN neutrinos were detected coming from the Large Magellanic Cloud when SN1987a underwent a core-collapse \cite{Kamiokande-II:1987idp,Bionta:1987qt,Alekseev:1988gp}, only the next SN observations will shed more conclusive light on the flavour composition of the neutrinos at the time of emission \cite{Capanema:2024hdm}.
     
    \item {\bf Neutrinoless Double Beta ($0\nu \beta \beta$) Decay:} Since $\epsilon$ is changing at a rate that alleviates some to all X-ray bounds, the importance of non-astrophysical experiments for placing bounds, or possibly detecting signals of sterile neutrinos, becomes highlighted. Apart from TRISTAN, one such prospect often stems from $0\nu \beta \beta$ decay, see for a review \cite{Rodejohann:2011mu}. Usually however, in the type-I seesaw with two sterile neutrinos and all neutrino masses below 100 MeV no $0\nu \beta \beta$ decay will be detectable \cite{deGouvea:2006gz,Girardi:2013zra}. Furthermore, when there are two sterile states with almost indistinguishable masses interference effects can appear and Ref.~\cite{Bolton:2019pcu} derives upper limits for the sum of squared active-sterile mixing $\abs{U_{e N_1}}^2 + \abs{U_{e N_2}}^2$ with only small mass splitting ratios between the steriles, as present in our model. They find usual $0\nu \beta \beta$ decay limits, which assume distinct masses, modified. In total, $0\nu \beta \beta$ bounds are not able to constrain our parameter regime for keV mass sterile neutrinos.
    
    \item {\bf 3.5 keV Line:} The observation of a 3.5 keV X-ray line sparked speculations whether it could originate from 7 keV sterile neutrino DM \cite{Bulbul:2014sua,Boyarsky:2014jta,Iakubovskyi:2015dna,Boyarsky:2014ska,Riemer-Sorensen:2014yda,Jeltema:2014qfa,Urban:2014yda,Dessert:2018qih,Boyarsky:2020hqb,Bhargava:2020fxr}. However, the line originated partly from objects sufficiently distant from Earth so that, in our model, the sterile neutrinos would not have been able to cause such a signal consistently. For instance, the 3.5 keV line has been observed in the Perseus and Coma galaxy clusters, located at distances $d>10^7$ light-years (see Table \ref{tab:Xray}). This implies that the photons observed on Earth today were emitted before the phase transition had commenced in the BM models. Consequently, the line cannot be attributed to sterile neutrino decay into a neutrino and a photon, as the active-sterile mixing at that time was $U_{eN} = 0$.

    \item \textbf{Dark Energy:} Since the scalar field in our scenario is typically light in the late cosmic history, it is natural to consider the possibility of identifying it with a dynamical dark energy field. Indeed, its dynamics in the broken phase is analogous to a thawing-quintessence \cite{Caldwell:2005tm, Amendola:2015ksp} scenario, where the Hubble friction freezes the field at a specific value until the Hubble rate becomes smaller than its mass, $H\lesssim\mu$ (see Ref.~\cite{Wetterich:2007kr} for the growing-neutrino quintessence scenario). However, our model requires $\mu(t_{LSB})\gg H(t_{LSB})$ since otherwise $\bar S$ would not be in the range \eqref{Srange} or the rate of change $\dot{\bar S}/\bar S \sim \bar{\mu}$ would be too small to evade X-ray constraints. Consequently, the scalar field cannot be frozen only by Hubble friction. Additionally, once the phase transition starts, the scalar field does not fulfil the usual slow-roll requirements, since its kinetic and potential energy are comparable in magnitude. Therefore, dark energy has to be explained by another independent mechanism. At any rate, the exponentially-growing expectation value of the scalar field induces a change in the overall cosmological expansion that might become observable in the relatively near future. 
\end{itemize}

\section{Discussion and Outlook}\label{sec:concl+outlook}

In this work, we have shown that it is possible to reconcile the keV mass range for DM sterile neutrinos with a successful neutrino mass generation by means of a very late cosmological phase transition that introduces a time dependence to the mixing between neutrino species. To this end, we considered a minimalist scenario involving only two RH neutrinos and an additional symmetry-breaking scalar field charged under a global $U(1)_N$ symmetry. The associated charge assignments ensure that active neutrinos gain masses only in the very late Universe, namely when the $U(1)_N$ symmetry is spontaneously broken. In contrast, the $\mathcal{O}$(keV) mass of the sterile DM neutrinos changes only minimally by the phase transition.
Given the tachyonic nature of the scalar field dynamics, the still-ongoing rolling phase leads to a time-dependent active-sterile mixing able to evade X-ray constraints, while reproducing the correct neutrino spectra. The degree to which the formerly excluded parameter space opens depends on the dynamics of the phase transition, which we choose to describe in a model-independent way in terms of a time-dependent mass term in the scalar potential, while two explicit realisations are presented in Appendix~\ref{app:ph_transition}.

The time-evolution of neutrino masses provides a clear way to test the predictions of our model. In fact, the vanishing of neutrino masses before the transition alleviates the tension between the DESI results and the oscillation data for neutrino masses. In other words, an apparent cosmological preference for massless neutrinos \textit{does not} require a deviation from the standard $\Lambda$CDM model and future galaxy surveys such as DESI \cite{DESI:2016fyo} or EUCLID \cite{Euclid:2024imf} are expected to fully exclude the minimal sum of neutrino masses from neutrino oscillation data. Another astrophysical probe could come from the study of the neutrinos produced in supernovae, since this will give us information about their masses at the time of production \cite{deGouvea:2022dtw,Capanema:2024hdm}. More directly, the predicted parameter space for active-sterile mixing will be probed by the TRISTAN extension to the KATRIN experiment \cite{KATRIN:2018oow} and potentially by HUNTER phase 3 \cite{Martoff:2021vxp}. A potential detection by TRISTAN or HUNTER would not only be in agreement with X-ray bounds but also provide a promising hint towards the here-presented idea. Finally, if the phase transition continues without the field being eventually stabilised by quartic or higher-order operators, there are a number of consequences that will become observable in the more distant future. The exponentially-growing expectation value of the scalar field will induce an increasing active-sterile mixing, therefore destabilising DM while increasing active neutrino masses and facilitating the probing of sterile neutrinos. Additionally, a change in the overall vacuum energy might become observable, eventually ending the $\Lambda$ domination. With dark matter and dark energy properties changing, the Universe as we know it is \emph{phasing out of darkness}.

\section{Acknowledgements}

We thank Salvador Centelles Chuliá for frequent and essential discussions during the preparation of this work. We are also grateful to Frederik Depta, Rasmi Hajjar, Joerg Jaeckel, Sophie Klett, Federica Pompa, Lucas Puetter, Werner Rodejohann, Manibrata Sen and Alexei Smirnov for useful inputs and discussions. GL (ORCID 0000-0002-4739-4946) is supported by a fellowship from ”la Caixa” Foundation (ID 100010434) with fellowship code LCF/BQ/DI21/11860024. JR is supported by a Ramón y Cajal contract of the Spanish Ministry of Science and Innovation with Ref.~RYC2020-028870-I. This work was supported by the project PID2022-139841NB-I00 of MICIU/AEI/10.13039/501100011033 and FEDER, UE.

\appendix

\section{Late Phase Transitions beyond the Effective Picture} \label{app:ph_transition}

Throughout the present work, we maintained a completely model-independent approach, focusing on pure phenomenological implications and assuming the late phase transition to be simply generated by a fast-changing mass that quickly turns tachyonic. In this Appendix, we put forward some prototypical embeddings of such a scenario, thus showcasing the wide range of applications of our results.   

\begin{itemize}
    \item \textbf{Non-Minimal Coupling to Curvature:} The desired time-dependence of the effective mass can be achieved by considering a non-minimal coupling $\xi \bar S^2 R$ of the scalar field to gravity, with $\xi$ a dimensionless coupling constant and $R$ the scalar curvature. The inclusion of such an interaction term is indeed essential for the self-consistency of the theory and the regularisation of the energy-momentum tensor on curved spacetimes \cite{Birrell:1982ix, Mukhanov:2007zz}. In this type of settings, the Ricci scalar acts essentially as a cosmic clock that can render the effective scalar field mass
    \begin{equation}
        \mu^2(t) \sim \xi R - \Bar{\mu}^2
    \end{equation}
    negative at a specific cosmic time (cf.~Refs.~\cite{Bettoni:2018utf,Bettoni:2018pbl,Bettoni:2019dcw,Bettoni:2021zhq,Laverda:2023uqv,Laverda:2024qjt} for comprehensive analyses in the context of second-order phase transitions, including applications to the Standard Model Higgs \cite{Laverda:2024qjt}, Ref.~\cite{Kierkla:2023uzo} for first-order phase transitions and Ref.~\cite{Bettoni:2021qfs} for a review). Indeed, for a FLRW background metric, the Ricci scalar $R=3(1-3w_{\rm eff}(t))H^2(t)$ is positive during expansion eras with effective equation of state $w_{\rm eff}<1/3$. This applies in particular to the inflationary stage, where the Hubble-induced field mass can easily outweigh the tachyonic contribution $-\bar \mu^2$ in the effective mass. During the post-inflationary radiation-dominated era, the stabilisation mechanism is no longer present but, due to the comparatively large Hubble scale, the Hubble friction term is still enough to freeze the field in the origin of the potential \cite{Alonso-Alvarez:2018tus}. In this phase, the tree-level decay $N \rightarrow S + \nu $ allowed by Eq.~\eqref{eq:Yukawa} becomes potentially relevant, since the scalar field is effectively massless. However, the lifetime associated to this decay channel is much larger than the age of the Universe if the effective Yukawa coupling is sufficiently suppressed by the new-physics scale $\Lambda_S$. Setting $\Lambda_S \gtrsim 10^{14} \text{ GeV}$ guarantees the stability of sterile neutrinos for $\mathcal{O}( \rm{keV})$ masses.
    
    The effective mass of the scalar field during matter domination remains positive until the Hubble function falls below the threshold $\Bar{\mu}^2 = 3 \xi H^2(t_{\rm LSB})$, which, for the scenarios considered in this paper, must occur only after matter-radiation equality at $H \sim 10^6 H_0$ but before today. Keeping the field frozen by Hubble friction before the onset of matter domination sets an upper bound on the bare mass of the scalar field, while the assumption of a currently ongoing phase transition, and therefore a non-frozen field at present time, results in a lower bound, leading to the requirement $H_0 \lesssim \Bar{\mu} \lesssim 10^6 H_0$. The exact timing of the transition depends, in a degenerate way, on the light tachyonic mass $\Bar{\mu}$ and the non-minimal coupling parameter $\xi$, being able to obtain very late phase transitions for simultaneously large $\xi$ and $\bar \mu$. Note that such large values for the non-minimal coupling parameter are a priori not excluded and help in stabilising the scalar field against quantum fluctuations during inflation. However, in the late Universe, the tiny cosmological Ricci scalar could be potentially modified by small-scale variations arising from local matter-energy distributions. Computing their impact requires an extrapolation of the spacetime metric from cosmological to astrophysical scales, a difficult task that goes beyond the reach of our present work.
\item  {\bf Pole Dark Energy:}
Pole dark energy scenarios provide another way of triggering the  necessary $S$ symmetry-breaking dynamics. In this type of setting, $S$ is coupled to an evolving quintessence field $\phi$ responsible for dark energy, whose kinetic term displays specific singular behaviours or ``poles" at certain field values. This leads to dramatic transformations in the physical variables describing the universe's expansion \cite{Galante:2014ifa,Artymowski:2016pjz}. As a specific realisation of this paradigm,  one could consider for instance a Lagrangian density 
 \begin{equation}\label{Salpha}
 { \frac{\mathcal{L}_{DE}}{\sqrt{-g}}} =  -\frac12  {\frac{(\partial \phi)^2}{\bigl(1-\frac{\phi^{2}}{\phi_*^2}\bigr)^{2}}} - U(\phi)-\frac12 (\partial \bar S)^2-V(S,\phi)  \,,
\end{equation}
with $U(\phi)$ a generic potential assumed only to be non-singular at the critical value $\phi =\phi_*$,
$$
U(\phi) = U_{+} + (\phi-\phi_*)\, \partial_{\phi}U |_{\phi = +\phi_*} +\ldots \,,
$$
and
\begin{equation}
V(S,\phi)=\frac{1}{2} \left({\phi}-\phi_*\right)^2 \bar S^2-\frac12\bar\mu^2 \bar S^2+V_{HO}(\bar S)
\end{equation}
a simple interaction potential among the dark energy field $\phi$ and $S$.
The origin of the kinetic pole in Eq.~\eqref{Salpha} can be understood within the framework of variable gravity \cite{Wetterich:2013jsa,Wetterich:2014gaa,Rubio:2017gty} or hyperbolic geometries \cite{Carrasco:2015uma}, naturally encountered in extended supergravities \cite{Kallosh:2017wnt,McDonough:2016der}. Transforming to a canonical variable $\varphi$, we obtain 
\begin{equation}
\frac{\partial\varphi}{\partial \phi}=\left(1-\frac{\phi^{2}}{\phi_*^2}\right)^{-1} \hspace{10mm}\longrightarrow \hspace{10mm}
\phi = \phi_*\, \tanh{\frac{\varphi}{\phi_*}} \,. 
\end{equation}
Note that for \(\varphi \rightarrow 0\), the two variables \(\varphi\) and \(\phi\) are identical. Therefore, provided that the crossover scale $\phi_*$ exceeds the tachyonic contribution $\bar \mu$, the mass of the scalar field $S$ in this regime remains positive definite, effectively locking it at the origin of the potential, with the $U(1)_N$ unbroken. This holds until the quintessence field reaches the vicinity of the boundary $\phi=\phi_*$, where
$\phi_*-\phi\, \simeq   2 e^{-2\varphi/\phi_* }$. Then, the primary distinction arises and all dark-energy interactions become quickly exponentially suppressed,
 \begin{equation}
\frac { \mathcal{L}_{\rm DE}}{ \sqrt{-g}} \simeq -\frac12 {(\partial_{\mu}\varphi)^{2}} -\frac12 {(\partial_{\mu} S)^{2}} -U_{+} +\frac12\bar\mu^2 \bar S^2  - V_{HO}(\bar S)  \,,
\end{equation} 
triggering the motion of scalar field expectation value $\bar S$ and resulting in the spontaneous breaking of the $U(1)_N$ . This provides a natural clock for triggering the phase transition, with the specific details depending on the form of $U(\phi)$ and $\phi_\ast$.
\end{itemize}

\section{Corrections to SM Processes and Fifth Force Constraints}
\label{app:fifthforce}

Adding a light scalar field with a changing expectation value can induce phenomenological issues. In the following, we address why our model is not plagued by existing experimental and observational bounds on time-varying fundamental constants, supernova energy loss and fifth-force constraints.

\begin{itemize}
    \item {\bf Time-Varying Constants:} By symmetry arguments, one should always be allowed to add $(S^c S)/\Lambda_S^2$ to any operator, since this is a singlet under $U(1)_{N}$. Although this type of insertion induces generically a time dependence in the mass of SM fields for $\bar S\neq 0$ of the type
\begin{equation}
    m_{f} \rightarrow m_{f} \left( 1 \pm \epsilon (t)^2\right)\,, \quad\quad m_{V}^2 \rightarrow m_{V}^2 \left( 1 \pm \epsilon (t)^2 \right)\,,
    \label{eq:couplingtofV}
\end{equation}
it is clear that the change in mass between e.g. the time of BBN/CMB ($\epsilon = 0$) and today ($\epsilon \sim 10^{-9}-10^{-11}$) is completely negligible, cf.~Eq.~\eqref{eq:vevRatio}. In addition to this, the $S$ field could directly couple to the electromagnetic field tensor,
\begin{equation}
    \mathcal{L} \supset - \frac{1}{4e^2} F^{\mu \nu} F_{\mu \nu} \left( 1 \pm \epsilon (t)^2 \right)\,,
    \label{eq:couplingtoF}
\end{equation}
making the fine-structure constant time-dependent, $\alpha \rightarrow {\alpha}/{(1 \mp \epsilon (t)^2)}$.  Like for the particle masses, the accumulated change since the start of the phase transition would be of the order $10^{-20}$, i.e. very suppressed and safely within the stringent bounds provided by the natural nuclear reactor Oklo at redshift $z = 0.14$, $\Delta \alpha / \alpha = (0.005 \pm 0.061) \times 10^{-6}$  \cite{Petrov:2005pu}, and atomic clocks at $z=0$, $ \text{d} \ln \alpha / \text{d} \ln a \leq (2.5 \pm 3.5) \times 10^{-9}$ \cite{Filzinger:2023zrs}. Indeed, for up to $\partial \epsilon / \partial (H_0 t) = 5 \times 10^{10} \, \epsilon$, the model is not in conflict with the Oklo limits and for both BM1 and BM2 the change is sufficiently small. 

\item {\bf Supernova Energy Loss:} Before the onset of the LSB, the mass of the scalar field $\mu(t)$ depends on the nature of the phase transition and is not specified here. Supernova energy loss bounds give bounds for $\mu \leq T^{SN}_{\text{core}} \sim 30$ MeV on the interactions that lead to Eq.~\eqref{eq:couplingtofV} and Eq.~\eqref{eq:couplingtoF} \cite{Stadnik:2015kia}. The limits set $\Lambda_S > \text{ a few } \times 10^3$ GeV, which will be respected by the constraint set from the following fifth-force considerations.

\item {\bf Fifth Force:} Light scalar fields can mediate a Yukawa--like force which modify the potential between two objects and can be constrained by fifth force searches. The searches have been used to place bounds on terms like $S \abs{H}^2$ \cite{Piazza:2010ye}, which however does not exist in our model, since $U(1)_{N}$ would not be conserved. In general, the fifth force constraints for terms quadratic in $S$ are less stringent, since the modification to the Yukawa potential is of the order $1/r^3$ \cite{Stadnik:2015kia}. In the Higgs portal, the quadratic term can appear without being suppressed, i.e. ${\mathcal{L} \supset \lambda_{SH} \abs{S}^2 \abs{H}^2}$. However, motivated experimentally by Higgs-To-Invisible decay searches, and from the requirement of non-thermalisation of $S$, the coupling $\lambda_{SH}$ is assumed to be negligible. The bounds on $\Lambda_S$ can differ if one allows different scales related to different couplings, but we will conservatively assume a global value. One bound stems from Refs.~\cite{Olive:2007aj} and \cite{Adelberger:2006dh}, that constrain the proton interaction parameter to be $\Lambda^\prime_p \gtrsim 2 \times 10^3$ GeV for $\mu \lesssim 10^{-4}$ eV. Therefore, as a conservative limit we set a lower cutoff limit $\Lambda_S \gtrsim 10^4 \text{ GeV}$, which has been used throughout this paper when evaluating viable parameter space.
\end{itemize}

\bibliographystyle{JHEP} 
\apptocmd{\thebibliography}{\justifying}{}{} 
\bibliography{bibliography}


\end{document}